\documentclass[pra,twocolumn,showpacs]{revtex4}
\usepackage{graphicx}
\usepackage{enumerate}
\usepackage{amsmath}
\usepackage{amssymb}
\usepackage[T1]{fontenc}

\newcommand{\e}{\mathrm{e}}

\newcommand{\p}{\partial}

\newcommand{\vid}[1]{\langle #1 \rangle}
\renewcommand{\Re}{\mathrm{Re}\,}
\renewcommand{\Im}{\mathrm{Im}\,}
\renewcommand{\i}{\mathrm{i}}

\begin{document}
\title{Schr\"{o}dinger-equation formalism for a dissipative quantum system}
\author{E. Anisimovas}\email{egidijus.anisimovas@ff.vu.lt}
\affiliation{Department of Theoretical Physics, Vilnius University,
Saul\.{e}tekio al.\ 9, LT-10222 Vilnius, Lithuania}
\author{A. Matulis}\email{amatulis@takas.lt}
\affiliation{Semiconductor Physics Institute, Go\v{s}tauto 11, LT-01108 Vilnius, Lithuania}

\begin{abstract}
We consider a model dissipative quantum-mechanical system realized by coupling
a quantum oscillator to a semi-infinite classical string which serves as a
means of energy transfer from the oscillator to the infinity and thus plays the
role of a dissipative element. The coupling between the two --- quantum and
classical --- parts of the compound system is treated in the spirit of the
mean-field approximation and justification of the validity of such an
approach is given. The equations of motion of the classical subsystem are solved
explicitly and an effective \textit{dissipative} Schr\"{o}dinger equation for the
quantum subsystem is obtained. The proposed formalism is illustrated by its
application to two basic problems: the decay of the quasi-stationary state and the
calculation of the nonlinear resonance line shape.
\end{abstract}


\date{\today}

\maketitle

\section{Introduction}

The rapid expansion of research frontiers into the nanometer and femtosecond regime
demands a careful consideration of dissipation as a way to counterbalance the energy
influx from the external fields in quantum mechanical systems.

The subject takes roots in the seminal work by Feynman and Vernon \cite{fey3} who
treated a quantum object coupled to an infinite collection of oscillators as a model
of a linear dissipative environment. By employing the path-integral techniques they were
able to eliminate the environment variables and arrived at a dissipative modification
of the Green's function of the quantum object. This approach was further used by a number
of authors, most notably by Caldeira and Leggett \cite{cal83}, who applied it to a
specific problem of dissipative effects in tunneling on a macroscopic scale. A detailed
account on the problem of dissipation in quantum-mechanical systems can be found in a
monograph by Weiss \cite{weiss06} while semiclassical approaches are also reviewed
in \cite{gross95}.

Generally, the main mode of attack uses the density-matrix formalism \cite{barnett05}
and the popular Lindblad technique \cite{gao97}. This approach is relevant for
\textit{open} quantum systems. In contrast, the considerably less arduous
Schr\"{o}dinger-equation formalism is regarded to be appropriate for \textit{isolated}
systems \cite{dekker77} that are described by a hermitian Hamiltonian.

Despite this fact, there were successful attempts to include dissipation directly
in the Schr\"{o}dinger equation. Most often they based on the de Broglie-Bohm
picture \cite{bohm87} and employed modifications to the quantum
potential \cite{spiller91}. These ideas were used by Kostin to derive a dissipative
nonlinear Schr\"{o}dinger equation \cite{kos72}, and Albrecht \cite{albr75} worked
out the general principles limiting the acceptable form of the nonlinear terms.

The problem of dissipation in quantum mechanics is closely connected to the use of
mixed quantum-classical treatments \cite{kapral99,wan02} whereby only a few degrees
of freedom of a complex system are treated quantum-mechanically while the remaining
ones are described classically. Although Egorov and coworkers \cite{egorov99} point
out that not always an accurate description is attained and so this approach must
be used with care, it is recognized as a method of choice when a significant reduction
of complexity is desired.

In this paper we present a derivation of a rather simple nonlinear Schr\"{o}dinger
equation that includes a dissipative term. This equation is obtained from a
consideration of a physical model: a quantum oscillator connected to an infinite
classical string that acts as a means to transfer energy away from the quantum
system (see also \cite{matulis06}). The obtained Schr\"{o}dinger-like equation is
illustrated by considering two specific examples. The first one, the decay of a
quasi-stationary state, is an auxiliary one and shows that the results obtained by
using our analysis do not contradict the ones derived by the usual and more complex
methods using the density-matrix formalism and a double set of variables. The second
example, the nonlinear resonance, is interesting on its own and demonstrates that
the quantum-mechanical nonlinear resonance is physically richer than its classical
analog.

Our paper is organized in the following way. In Sect.~II, the model system is introduced
and the dissipative Schr\"{o}dinger-like equation for the quantum system is derived. In
Sect.~III, the obtained equation is simplified employing the rotating-wave approximation.
In the next two Sections, some illustrations of the developed technique are presented:
in Sect.~IV we consider the decay of the quasi-stationary state, and Sect.~V is devoted
to the non-linear resonance. Finally, our conclusions are formulated in Sect.~VI.

\section{Interaction of the quantum oscillator with a classical chain}

\begin{figure}[h]
\begin{center}\leavevmode
\includegraphics[width=80mm]{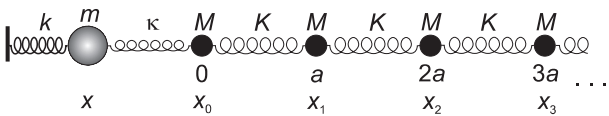}
\caption{The layout of the model system.}
\label{fig1}
\end{center}
\end{figure}

Let us begin by introducing our model of weakly coupled quantum and classical
subsystems. The quantum system is modeled as a harmonic oscillator of mass $m$
and spring constant $k$, described by the quantum-mechanical position and
momentum operators $\hat{x}$ and $\hat{p}$. The role of the classical system is
played by a semi-infinite classical string which, for the sake of convenience, is
discretized and represented as a chain of identical balls of mass $M$ interconnected
by springs of equilibrium length $a$ and spring constant $K$. The two subsystems
are linked by a different spring (spring constant $\kappa$) connecting the quantum
oscillator to the ball sitting at the end of the chain as depicted in Fig.~\ref{fig1}.
The corresponding Hamiltonian reads
\begin{equation}\label{ham}
\begin{split}
  H =& \frac{1}{2m}\hat{p}^2 + \frac{k}{2}\hat{x}^2
  + \frac{\kappa}{2}(\hat{x} - x_0)^2 \\
  &+ \sum_{n=0}^{\infty}
  \left[\frac{1}{2M}p_n^2 + \frac{K}{2}(x_n - x_{n+1})^2\right],
\end{split}
\end{equation}
with $x_n$ denoting the coordinate of the $n$-th ball measured from its equilibrium
position, and $p_n$ the respective momentum.

Following \cite{boucher88}, we assume that the quantum oscillator is described by
its own wave function $\Psi(x,t)$ which solves the Schr\"{o}dinger equation
\begin{equation}\label{schred}
  \i\hbar\frac{\p}{\p t}\Psi(x,t) = H\Psi(x,t)
\end{equation}
with the above Hamiltonian (\ref{ham}) while the dynamical variables of the balls in
the chain obey the classical Hamilton equations
\begin{equation}\label{newton}
  \dot{x}_n = \frac{\p}{\p p_n}\vid{H}, \quad
  \dot{p}_n = -\frac{\p}{\p x_n}\vid{H},
\end{equation}
with the Hamiltonian operator replaced by its quantum-mechanical average
over the state of the quantum oscillator
\begin{equation}\label{twohvid}
  \vid{H} = \int_{-\infty}^{\infty}dx\Psi^*(x,t)H\Psi(x,t).
\end{equation}
In this way, quantum variables and operators do not enter the equations for
the classical degrees of freedom and one obtains a consistent description.
This averaging constitutes the main assumption of the quasiclassical approximation
and was used earlier to treat the coupling of classical and quantum degrees
of freedom \cite{boucher88}. We give some further justification of this crucial step
below.

Eqs.~(\ref{newton}) lead to the standard classical equations for the chain
\begin{subequations}\label{chlg}
\begin{eqnarray}
\label{chlg1}
  M\ddot{x}_0 + K(x_0 - x_1) + \kappa x_0 &=& \kappa\vid{x}, \\
\label{chlg2}
  M\ddot{x}_n - K(x_{n-1} - 2x_n + x_{n+1}) &=& 0, \; n \geqslant 1.
\end{eqnarray}
\end{subequations}
The second of these equations (\ref{chlg2}) is readily solved by the Fourier
transform
\begin{equation}\label{furjeint}
  x_n = \frac{1}{2\pi}\int_{-\infty}^{\infty}dq\,\e^{\i(qan - \omega t)}f(q),
\end{equation}
producing the standard dispersion law
\begin{equation}\label{displaw}
  \omega^2 = \frac{4K}{M}\sin^2(qa/2).
\end{equation}
From here on we restrict our consideration to the long-wavelength approximation
($qa \ll 1$) thus neglecting the discretization and effectively returning to a
continuous-string model for the classical subsystem. In this limit, we have a
linear dispersion law
\begin{equation}\label{ldl}
  \omega = vq
\end{equation}
with the wave-propagation velocity $v = \Omega a$ and $\Omega = \sqrt{K/M}$.

Inserting the Fourier transform (\ref{furjeint}) into the boundary condition
(\ref{chlg1}) and using the long-wavelength approximation we are able to simplify
its left-hand side to
\begin{equation}\label{ks}
\begin{split}
& \frac{1}{2\pi}\int_{-\infty}^{\infty}dq\,\e^{-\i\omega t}
  \left\{K\left(1 - \e^{\i qa}\right) - Mv^2q^2 + \kappa\right\}f(q) \\
&  \approx \frac{1}{2\pi}\int_{-\infty}^{\infty}dq\,
  \e^{\i vqt}(-i Kqa + \kappa)f(q) \\
&  = \frac{Ka}{v}\dot{x} + \kappa x_0.
\end{split}
\end{equation}
The resulting equation of motion for the first ball coordinate $x_0$
\begin{equation}\label{ks2}
 \frac{Ka}{v}\dot{x}_0 + \kappa x_0 = \kappa\vid{x}
\end{equation}
may be interpreted as the equation of motion for a dissipative classical mode.
Eq.~(\ref{ks2}) together with the Schr\"{o}dinger equation (\ref{schred}) for
the quantum variables constitute the complete equation set describing the
behavior of the dissipative quantum system.

Now we turn to the examination of the validity of the employed quasiclassical
approximation. A smooth way to proceed is to introduce the dimensionless variables
by means of scaling
\begin{equation}\label{twoscale}
\begin{split}
  \hat{x} \to &l\hat{x}, \quad \hat{p} \to \frac{\hbar}{l}\hat{p}, \quad
  x_0 \to l_0x_0, \quad t \to \omega_0^{-1}t, \\
  &\omega_0 = \sqrt{\frac{k}{m}}, \quad
  l = \sqrt{\frac{\hbar}{m\omega_0}}, \quad
  l_0 = \sqrt{\frac{\hbar}{M\Omega}}.
\end{split}
\end{equation}
Here $l$ and $l_0$
are, respectively, the characteristic length scales of the quantum and classical
subsystems, and $\omega_0$ stands for the frequency of the quantum oscillator.
The scaling enables us to rewrite the obtained set of equations as
\begin{subequations}\label{setfin}
\begin{eqnarray}
\label{setfin1}
&&  \i\frac{\p}{\p t}\Psi(x,t) = H\Psi(x,t), \\
\label{setfin2}
&&  H = \frac{1}{2}\hat{p}^2 + \frac{1}{2}\hat{x}^2 - \lambda\hat{x}x_0, \\
\label{setfin3}
&&  \dot{x}_0 + \lambda\xi x_0 = \lambda\vid{x},
\end{eqnarray}
\end{subequations}
with
\begin{equation}\label{interact}
  \lambda = \frac{\kappa}{k_0}\sqrt{\frac{\omega_0m}{\Omega M}}, \quad
  \xi = \frac{l_0}{l}.
\end{equation}
Note that we excluded the terms which do not depend on the operators $\hat{x}$
and $\hat{p}$ from the quantum Hamiltonian (\ref{setfin2}) as they can always
be absorbed into the wave-function phase. The obtained equations describe the
interaction of a quantum-mechanical system with a single dissipative classical
mode $x_0$. In fact, the derivation does not depend on the assumption that
the quantum system is a harmonic oscillator. Thus, the equations are general
and can be applied to any quantum system with dissipation. For instance,
considering an \textit{anharmonic} oscillator driven by an external time-periodic
force (we treat this case as an illustration) the above Hamiltonian is
supplemented with the following additional terms
\begin{equation}\label{hamadd}
  \Delta H = -\hat{x}f\cos(\omega t) + \alpha\hat{x}^4,
\end{equation}
with $f$ expressing the force amplitude and $\alpha$ being the anharmonicity
coefficient.

In order to have an appreciable (that is, not negligibly small) interaction
between the quantum and classical subsystems one has to assume the characteristic
frequencies $\omega_0$ and $\Omega$ to be of the same order of magnitude. Under
this assumption the quantum oscillator will be able to emit ``phonons'' into the
string. Furthermore, the coupling constant $\kappa$ representing the
interaction of quantum oscillator with the chain should not exceed the
constant $k$ describing the oscillator potential itself. Otherwise, the
oscillator cannot be considered as a separate entity. In view of the above
constraints, the dimensionless coupling constants $\lambda$ and $\xi$ depend
essentially only on the adiabatic parameter $m / M$.

Thus, in the case of a light quantum oscillator interacting with a chain of heavy
balls the above adiabatic parameter is small and so are the coupling constants.
The smallness of the coupling constant ensures
a weak perturbation of the chain. The chain under consideration may be
treated as a collection of noninteracting harmonic modes. Meanwhile, the
quantum-mechanical description of a harmonic mode gives the same result as
the classical one. Therefore, one may conclude that the smallness of the
adiabatic parameter $m / M$ guarantees the validity of the applied quasiclassical
approximation for the description of the interaction between the quantum and
classical subsystems. A similar argument was also presented in the development of
the mixed quantum-classical dynamics \cite{kapral99}.

Alternatively, one may argue that due the smallness of the adiabatic parameter the
characteristic length of the wave functions of heavy balls $l_0$ is much smaller
than the characteristic length $l$ of the wave function of the oscillator.
Consequently, the balls may be considered to be following their well-defined
classical trajectories whereas the oscillator has to be described quantum-mechanically.

\section{Rotating-wave approximation}

Let us now proceed to some illustrations of the developed formalism based on
a weakly nonlinear oscillator. Namely, we assume that the Hamiltonian of the
quantum subsystem consists of a sum of Eq.~(\ref{setfin2}) and Eq.~(\ref{hamadd}).
Besides, we restrict the treatment to the case of a weak nonlinearity, that is,
we assume that the anharmonicity $\alpha$, the force amplitude $f$, and the coupling
constants $\lambda$ and $\xi$ are small quantities compared to the unity. We also
restrict the frequency of the driving field to the immediate vicinity of the
resonant frequency of the harmonic oscillator which is unity in the dimensionless
units. Thus we are able to treat the frequency deviation from the resonance as yet
another small parameter $\eta = \omega - 1$.

We expand the sought wave function into the series of harmonic-oscillator
eigenfunctions
\begin{equation}\label{series}
  \Psi(x,t) = \sum_{n=0}^{\infty}a_n(t)\e^{-\i\omega E_n t}\psi_n(x).
\end{equation}
Here, the dimensionless energies and the corresponding eigenfunctions read
\begin{equation}\label{eigen}
  E_n = n + \frac{1}{2}, \quad \psi_n(x)
  = \frac{1}{\sqrt{2^nn!\sqrt{\pi}}}\e^{-x^2/2}H_n(x),
\end{equation}
and the symbol $H_n(x)$ stands for the $n$-th Hermite polynomial. Note that
instead of the usual time dependence $\sim\e^{-\i E_n t}$ of a harmonic-oscillator
eigenfunction Eq.~(\ref{series}) features a slightly modified exponential
factor $\e^{-\i \omega E_n t}$ that absorbs a part of the time dependence
of the expansion coefficient $a_n(t)$.

Using the above expansion of the wave function (\ref{series}) the coordinate
average can be presented as
\begin{equation}\label{aver}
\begin{split}
  \vid{x} &= \sum_{n,m=0}^{\infty}\e^{\i\omega(E_n - E_m)t}a_n^*a_m\vid{n|x|m} \\
  &= \sqrt{2}\,\Re\left(\e^{\i\omega t}a\right)
\end{split}
\end{equation}
where
\begin{equation}\label{polar}
  a = \sum_{n=0}^{\infty}a_na_{n+1}^*\sqrt{n+1}
\end{equation}
is the usual polarization of the quantum oscillator.

Inserting the coordinate average (\ref{aver}) into Eq.~(\ref{setfin3}),
and solving it to the leading term in the small coupling constants we obtain
the following expression for the coordinate of the first ball
\begin{equation}\label{fbex}
  x_0 = \sqrt{2}\,\lambda\int dt\,\Re\left(\e^{\i\omega t}a\right)
  = \sqrt{2}\lambda\,\Im\left(\e^{\i\omega t}a\right).
\end{equation}

Next, inserting the obtained expression (\ref{fbex}) and the wave function
(\ref{series}) into Eq.~(\ref{setfin1}), projecting this equation onto the $n$-th
state, and neglecting all oscillating terms (this amounts to the usual
rotating-wave approximation) we arrive at the final set of equations for the
time evolution of the wave function expansion coefficients
\begin{eqnarray}\label{fsa}
  &&  \i\dot{a}_n = -\left\{\eta E_n - \alpha
  \vid{n|x^4|n}\right\}a_n \nonumber \\
  &&  -\sqrt{n}(F + \i\gamma a^*)a_{n-1}
  -\sqrt{n+1}(F - \i\gamma a)a_{n+1}.\phantom{mm}
\end{eqnarray}
Here, for the sake of simpler notation we denoted
\begin{equation}\label{not}
  F = f/2\sqrt{2}, \quad \gamma = \lambda^2/2.
\end{equation}

The obtained set of equations (\ref{fsa}) was solved in the two- and three-level
approximations, and the results are presented in the following sections.

\section{Decay of the quasi-stationary state}

In order to illustrate the validity of the proposed method we consider a
well-known problem of the decay of a quasi-stationary state.

For this purpose, we assume that the quantum system has only two states and there
is no anharmonicity and no external force ($F = \eta = \alpha = 0$). At the
beginning the system is prepared in the upper level, that is, the boundary condition
reads
\begin{equation}\label{twop}
  t = -\infty, \quad a_0 = 0, \quad a_1 = 1.
\end{equation}

\begin{figure}[ht]
\begin{center}\leavevmode
\includegraphics[width=80mm]{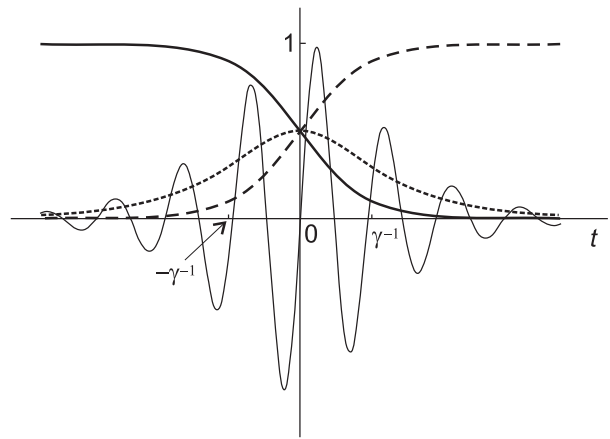}
\caption{Decay of quasi-stationary state: $|a_1|^2$ -- thick solid curve,
$|a_0|^2$ -- dashed curve, $a$ -- dotted curve, and $x_0$ -- thin solid curve.}
\label{fig2}
\end{center}
\end{figure}

According to Eq.~(\ref{fsa}), the behavior of this simple system is described
by the following two equations
\begin{equation}
  \dot{a}_0 = \gamma aa_1, \quad
  \dot{a}_1 = -\gamma a^*a_0, \quad \textrm{with }a = a_0 a_1^*.
\end{equation}
These equations are readily solved analytically, and the solution
reads
\begin{equation}\label{twodecsol}
  |a_0|^2 = \frac{1}{\e^{-2\gamma t} + 1}, \quad
  |a_1|^2 = \frac{1}{\e^{2\gamma t} + 1}.
\end{equation}
These filling factors are shown in Fig.~\ref{fig2} by the thick solid (excited level)
and dashed (ground level) lines.
We see that the quantum transition takes place during a finite period of
time proportional to $\gamma^{-1}$, the reciprocal friction coefficient.
The dotted curve shows the oscillator polarization
\begin{equation}\label{pol}
  a = \frac{1}{2\cosh(\gamma t)}
\end{equation}
induced during the quantum transition, and the thin solid line shows the
classical coordinate of first ball in the chain $x_0(t)$ which is calculated
by means of expression (\ref{fbex}). As a matter of fact, this curve represents
the coordinate of \textit{any} ball in the chain because according to
Eq.~(\ref{furjeint}) the coordinate of the $n$-th ball can be expressed as
\begin{equation}\label{xncoord}
  x_n(t) = x_0(t - an/v).
\end{equation}
Therefore, during the decay of the quantum state a string excitation of the shape
shown by thin solid line is emitted and travels away from the quantum oscillator with
a constant velocity $v$. It can be interpreted as the classical analog of an emitted
phonon. Naturally, the total energy carried by the string excitation is equal to the
energy difference between the two quantum levels. Consequently, we may conclude that
the quantum subsystem provides the energy quantization in the classical one.

It is worth mentioning that taking into account the norm conservation
($|a_0|^2 + |a_1|^2 = 1$) Eqs.~(\ref{fsa}) can be transformed into
another set of equations governing the behavior of three real variables:
the inversion $I = |a_1|^2 - |a_0|^2$ and the (complex) polarization $a$.
The linearized version of these equations in the vicinity of the stationary
state $n=0$ ($I=-1$) reads
\begin{equation}\label{dmlin}
  \dot{I} = -2\gamma I, \quad
  \dot{a} = -\gamma a.
\end{equation}
It is remarkable that these equations coincide with those considered in the
density matrix formalism \cite{wod79}, moreover, from our Eq.~(\ref{dmlin}) we are
able to extract the relaxation times: the longitudinal relaxation time is
$T_1=1/2\gamma$ and the perpendicular one is $T_2=1/\gamma$. We see that these
relaxation times obey the
relation $T_2=2T_1$ inherent to the pure state when no ensemble average is carried
out and this fact confirms the adequacy of our attempt to describe a dissipative
quantum system in the Schr\"{o}dinger-equation formalism.

Two more comments are in order. First, according to Eqs.~(\ref{twop}) we used the
boundary condition at $t=-\infty$. This is related to the classical description
of the chain excitations when there is no spontaneous emission. When solving the
equations numerically one has to add some initial vibration field fluctuation in
order to force the quantum transition.

Another point is related to the Fourier transform of the coordinate of the first ball
\begin{equation}\label{ftfb}
  x_0(\omega) = \int_{-\infty}^{\infty}d\omega\, e^{\i\omega t}x_0(t)
  \sim \frac{\pi/2\gamma}{\cosh(\eta\pi/2\gamma)}
\end{equation}
which gives us the shape of the emission line. We see that close to the resonant
frequency, when $|\eta| \ll 1$, the line shape is Lorentzian coinciding with the result
following from a simple calculation based on the Fermi's golden rule. However, further
away from the resonance, the line shape deviates from the Lorentzian and features
exponential tails.

\section{Resonant power absorption}

The following problem is that of the resonant power absorbtion which we calculated
solving the equation set (\ref{fsa}) using two-level
\begin{subequations}\label{reztwo}
\begin{eqnarray}
  \dot{a}_0 &=& \i\frac{\eta}{2}a_0 + \i (F - \i\gamma a)a_1, \\
  \dot{a}_1 &=& \i\frac{3\eta}{2}a_1 + \i (F + \i\gamma a^*)a_0, \\
\label{dulg2}
  a &=& a_0a_1^*.
\end{eqnarray}
\end{subequations}
and three-level
\begin{subequations}\label{rezthree}
\begin{eqnarray}
  \dot{a}_0 &=& \i\left(\eta/2 - \alpha\right)a_0
  + \i(F - \i\gamma a)a_1, \\
  \dot{a}_1 &=& \i\left(3\eta/2 - 5\alpha\right)a_1 \nonumber \\
  && + \i(F + \i\gamma a^*)a_0 + \sqrt{2}\,\i(F - \i\gamma a)a_2, \\
  \dot{a}_2 &=& \i\left(5\eta/2 - 13\alpha\right)a_1
  + \sqrt{2}\,\i(F + \i\gamma a^*)a_1, \\
\label{tryslg2}
  a &=& a_0a_1^* + \sqrt{2}\,a_1a_2^*.
\end{eqnarray}
\end{subequations}
approximations. Note that we excluded a nonlinear term from the equations of the
two-level approximation because in this case it leads only to an inessential shift
of the resonant frequency. The power absorption was calculated by averaging the
instantaneous power over the period of the external force, that is,
\begin{equation}\label{abspower}
  P = -2F\Im a.
\end{equation}

\begin{figure}[ht]
\begin{center}
\includegraphics[width=80mm]{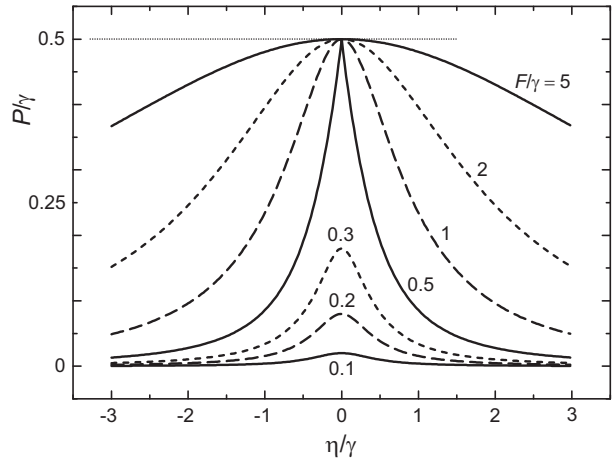}
\caption{Power absorption in the two-level approximation. The numbers on curves
indicate the values of the external force amplitude $F/\gamma$.}
\label{fig3}
\end{center}
\end{figure}

The power absorption in the two-level approximation as a function of the deviation
$\eta$ is shown in Fig.~\ref{fig3} for various force amplitudes. Note that
Eqs.~(\ref{reztwo}) and (\ref{rezthree}) depend linearly on their parameters,
therefore, by using an appropriate scaling of the time we may reduce the
number of parameters by one. Therefore, we choose to express the quantities $F$,
$P$, $\eta$ and $\alpha$ in $\gamma$ units.
We observe the following typical behavior: For a weak force ($F_0 < 0.5\gamma$)
the shape of the resonance resembles a Lorentzian, and its peak value is increasing
with increasing force. When the force exceeds the critical value $F_0 = 0.5\gamma$
the peak stops growing and is flattened at the top due to the saturation at the
critical power absorption value $P_0 = 0.5\gamma$ indicated by the thin dotted line.

\begin{figure}[ht]
\begin{center}
\includegraphics[width=80mm]{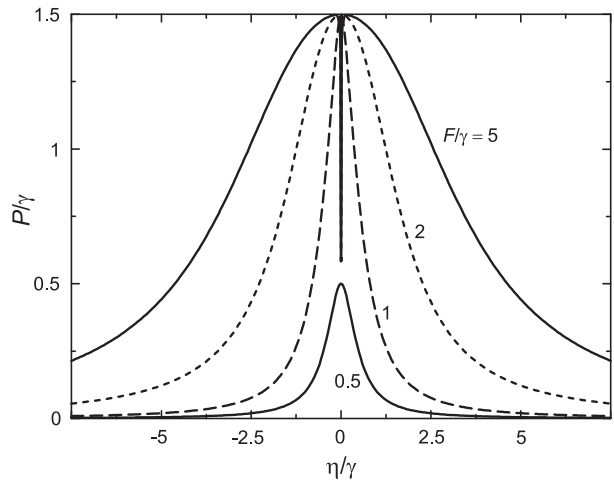}
\caption{Power absorption in the three-level approximation in the case of the
linear oscillator. The numbers on the curves indicate the values of the external force
amplitude $F / \gamma$.}
\label{fig4}
\end{center}
\end{figure}

In the next two Figures the results for power absorption calculated in the three-level
approximation are shown. In Fig.~\ref{fig4} we plot the power absorption in a
linear oscillator ($\alpha = 0$).

The general shape of the absorption curves closely resembles those obtained in the
two-level approximation, except that now the saturation sets in at a larger
force value $F_0 = \gamma$, and the saturated power absorption $P_0 = 1.5\gamma$
is larger as well.

A more essential difference is seen in the uppermost curve corresponding to the force
whose amplitude ($F = 5\gamma$) strongly exceeds the critical value. Here we see
a narrow gap in the absorption which appears close to the resonant frequency. It
resembles an analogous gap that appears in the power absorption calculated by means
of the density matrix equation technique \cite{gap} and is caused by the interaction
of coherent light with an inhomogeneously broadened resonance line.

\begin{figure}[ht]
\begin{center}
\includegraphics[width=80mm]{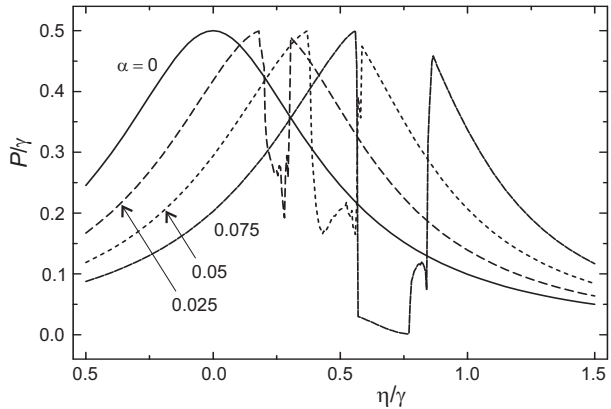}
\caption{Power absorption in the three-level approximation in the
case of an anharmonic oscillator obtained for the parameter $F =
0.5\gamma$. The curves are numbered by the values of
the anharmonicity coefficient $\alpha/\gamma$.} \label{fig5}
\end{center}
\end{figure}
In Fig.~\ref{fig5} the results for the nonlinear oscillator are presented for
a set of values of the anharmonicity coefficient and the force amplitude
corresponding to the critical value of the two-level approximation,
i.~e.\ $F = 0.5\gamma$. We see that for the
linear-oscillator case ($\alpha = 0$) the absorption demonstrates the expected
Lorentzian behavior. When the anharmonicity is incremented the resonant peak moves
to the right, becomes asymmetric and the saturation manifests itself as a
widening absorption gap close to the shifted resonance frequency.

The absorption gap resembles the gap in nonlinear cyclotron resonance
\cite{pyragas85} which was described using the balance equations. In that work
the gap was related to the dynamical chaos appearing in the \textit{classical}
nonlinear equations of motion. Quantum chaos is a subject of great interest as
well (see \cite{gutzw98} for references), and is largely concerned with the
quantum signatures of classically chaotic systems. Such signatures are often
sought by comparing the wave-function patterns corresponding to various
eigenstates with the classical trajectories \cite{heller93}.

\begin{figure}[ht]
\begin{center}
\includegraphics[width=80mm]{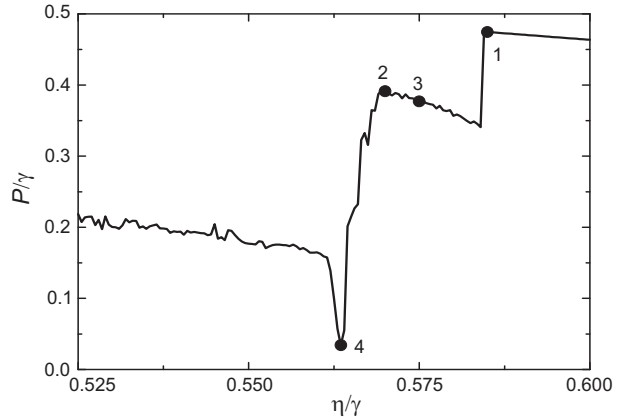}
\caption{Power absorption gap for the parameter values
$F = 0.5\gamma$ and $\alpha = 0.25\gamma$.}
\label{fig6}
\end{center}
\end{figure}

\begin{figure}[ht]
\begin{center}
\includegraphics[width=80mm]{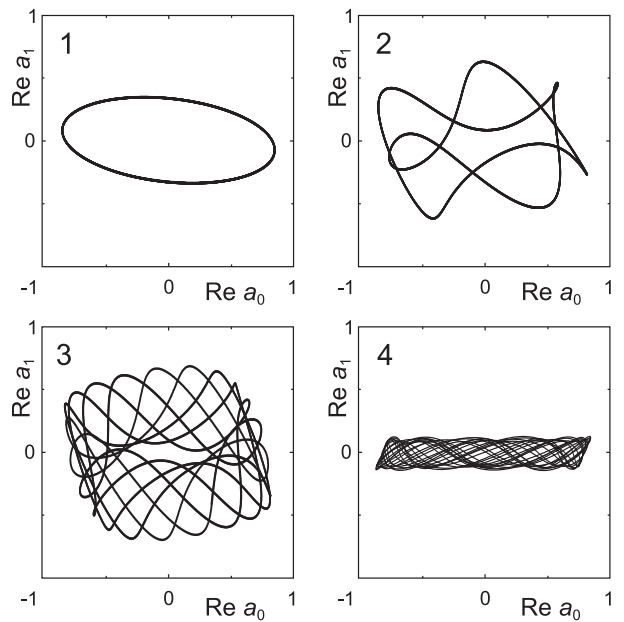}
\caption{Two-dimensional projection of the phase space. The numbers on the panels
correspond to the frequencies indicated by dots in Fig.~\ref{fig6}.}
\label{fig7}
\end{center}
\end{figure}

It follows from our quasiclassical consideration, Eqs.~(\ref{fsa}),
that the behavior of the quantum system can also be visualized and interpreted
in terms of phase-space trajectories obtained by treating the wave function
expansion coefficients $a_n$ as generalized coordinates. In order to illustrate
this possibility, we looked at the behavior of the quantum anharmonic oscillator
in the absorption gap more carefully. In Fig.~\ref{fig6} the right-hand side of
this absorption gap for the parameter values $F = 0.5\gamma$ and
$\alpha = 0.25\gamma$ is shown in more detail. The simplest way to distinguish
various types of dynamic behavior is to plot the phase portrait of the system.
As there are three complex variables constrained by a single normalization
condition the phase space is five-dimensional.
Therefore, we restrict ourselves to a projection of the phase space onto a
two-dimensional plane $(\Re a_0, \Re a_1)$ shown in Fig.~\ref{fig7}. The panel
numbers correspond to the frequencies indicated by dots in Fig.~\ref{fig6}.
We see that on the regular flank of the resonance curve (point 1) the oscillator
demonstrates a rather simple behavior --- the limit cycle. As a matter of fact, in
this case all three coefficients $a_n$ oscillate with the same frequency, namely,
$a_n(t)=g_n\exp(\i\zeta t)$. Therefore, in this case the power absorption problem
can be simplified essentially and transformed into a nonlinear eigenvalue problem
defining the frequency $\zeta$, and the stationary coefficients $g_n$ are simply
related to the filling factors of the harmonic oscillator levels.

The behavior of the anharmonic oscillator in the absorption gap (points 2, 3, 4)
is quite different. Here, the coefficients $a_n$ oscillate with multiple frequencies,
moreover, the oscillator can demonstrate both periodic behavior (point 2) and
non-periodic behavior (points 3, 4). This non-periodicity is in fact
responsible for the gap in the power absorption as it causes the loss of coherence
between the driving force and the system response. This is the so-called quasiperiodic
behavior but the appearance of a strange attractor is also expectable at larger
values of the anharmonicity coefficient $\alpha$.

\section{Conclusions}

In conclusion, we propose a Schr\"{o}dinger-like equation for the description
of a dissipative quantum-mechanical system. This equation is derived on the basis
of a quasiclassical treatment of coupled quantum and classical subsystems and
justified by the difference of masses of the two subsystems which leads to the
smallness of adiabatic parameters. The considered decay of a quasi-stationary state
demonstrates that the proposed description gives identical results to those obtained
from the density-matrix approach when only homogeneous broadening is taken into
account. The model calculation of the nonlinear absorption indicates that many
classical phenomena --- such as the asymmetry of the resonance shape and
absorption dips --- brought about by the loss of coherence between the driving
force and the response may be featured by a quantum system as well. We believe that
the proposed Schr\"{o}dinger-equation formalism with dissipation may
find application to quantum nanostructures and quantum chaos.

\begin{acknowledgments}
We thank Prof.~K.~Pyragas and J.~Ruseckas for helpful discussions. This work was
supported by the Lithuanian Science and Studies Foundation under grant No.~T-05/18.
\end{acknowledgments}

\end{document}